\documentclass[aps,pra,showpacs,showkeys]{revtex4} 
\usepackage{graphicx}

\def\ket#1{|\,#1\,\rangle}

\def\opone{\leavevmode\hbox{\small1\kern-3.8pt\normalsize1}}

\newcommand{\beq}{\begin{equation}}
\newcommand{\eeq}{\end{equation}}
\newcommand{\ba}{\begin{eqnarray}}
\newcommand{\ea}{\end{eqnarray}}
\newcommand{\bea}{\begin{eqnarray}}
\newcommand{\eea}{\end{eqnarray}}
\newcommand{\bma}{\begin{subequations}}
\newcommand{\ema}{\end{subequations}}
\newcommand{\bwt}{\begin{widetext}}
\newcommand{\ewt}{\end{widetext}}

\begin{document}

\title{Off-resonant Raman echo quantum memory on atoms with natural inhomogeneous broadening in optical QED cavity}

\author{S. A. Moiseev}
\email[]{samoi@yandex.ru}
\affiliation{Kazan Physical-Technical Institute of the Russian Academy of Sciences, Russia}
\pacs{ 03.67.-a, 03.67.Hk, 42.50.Md, 42.50.Ex.}
\keywords{optical quantum memory, photon echo, off-resonant Raman transitions, natural inhomogeneous broadening, optimal optical QED cavity, two impedance matching conditions.}

\begin{abstract}
A new scheme of photon echo based quantum memory in the optimal optical QED cavity with off-resonant Raman atomic transition is proposed.
The scheme employs the atomic ensembles characterized by an optically thin resonant transition and natural inhomogeneous broadening of the resonant line composed of the arbitrary narrow homogeneous spectral components. The scheme provides robust and quite simple coherent control of the light-atoms dynamics that can be implemented by using an existing optical technique and opens a practical way for realization of the efficient long-lived  multi-mode optical quantum memory.
\end{abstract}

\date{\today}

\maketitle

\section{Introduction}

Realization of efficient multi-mode quantum memory (QM) is a topical problem of quantum optics and quantum information science. There are a number of proposals which have provided a large progress in the last decade \cite{Lvovsky2009,Hammerer2010,Simon2010,Tittel2010}, however a simultaneous realization of main requirements such as almost ideal quantum efficiency and fidelity, long-lived and multi-mode capacity of QM is still the subject of intensive investigations \cite{JPhysB2012}. Photon echo based QM \cite{Moiseev2001} is one of the promising techniques where most successful experimental results have demonstrated for the efficient quantum storage of multi-mode light fields in a free propagation scheme with controlled reversible inhomogeneous broadening (CRIB) realized via switching of the external electric (or magnetic) field gradients \cite{Hedges2010,Hosseini2011} (known as a gradient echo memory (GEM). However the potentially acceptable properties of this technique have their limitations in a maximum value of controlled inhomogeneous broadening (IB) and in the realization of ultimately narrow homogeneous broadening and therefore these two factors confine the multi-mode capacity, atomic decoherence and optical depth on the resonant atomic transition. In order to resolve these problems, the new approaches have been proposed later such as the photon echo technique QM on the IB line composed of the periodic atomic frequency combs (AFC-protocol) \cite{Riedmatten2008} which has been successfully demonstrated for the broadband \cite{Clausen2011,Saglamyurek2011} and multi-mode storage \cite{Usmani2010,Bonarota2011} (see also the modified AFC scheme \cite{Moiseev2012} providing theoretically almost 100 \%  quantum efficiency for the broadband light fields and new scheme of QM based on broadband slow light in AFC-media \cite{Bonarota2012}). However, a sufficiently high quantum efficiency can be realized only for rather narrow AFCs which remains a serious experimental problem. There are other proposals of the photon/spin echo based QMs, in particular using the natural IBs \cite{Moiseev2011, McAuslan2011,Damon2011,BHam2012}. However so far all the approaches still have physical problems in experimental realizations of the practically significant physical properties.

Recently we have proposed a photon echo QM in the optimal resonant cavity with CRIB technique for rephasing of atomic coherence and demonstrated its effective integration into the quantum computer scheme \cite{Moiseev2010}.  The QED cavity approach provides a perfect time reversal retrieval of the multi-mode light fields for optically thin atomic system if an ideal CRIB procedure \cite{Moiseev2001, Moiseev2004,Kraus2006} can be successfully realized. Similarly the AFC protocol has been proposed for QED cavity \cite{Afzelius2010}, meanwhile it can not provide a time reversal dynamics. The recent experiment \cite{Sabooni2012} has confirmed the  possibility of photon echo QM with optically thin media. Perfect CRIB procedure in GEM technique can be implemented on the atomic ensembles provided that resonant linewidth is reduced to narrow homogeneous broadening for strong elimination of the atomic decoherence but it decreases the effective optical depth and can lead to some experimental difficulties. For example, realization of CRIB procedure in QED cavity can  disturb the cavity tuning and resonant interaction of the cavity mode with atomic ensemble. Thus it is highly desirable to find a simpler experimental solution to overcome the  physical problems related to the rephasing of macroscopic coherence in IB atomic systems. In this paper we propose a new simple scheme for highly efficient multi-mode optical QM which resolves basic problems by using well-known experimental technique. For this purpose  we combine the specific advantages of the photon echo QM in optical QED-cavity with off-resonant Raman type of atomic transition \cite{Moiseev2011b, Nunn2008, Gouet2009, EMoiseev2013} in one scheme. Thus the proposed scheme almost ideally works with natural IB line characterized by ultra-narrow homogeneous isochromatic components. Also the used optical QED-cavity opens a convenient way for delicate rephasing of the atomic coherence excited in IB system without significant excitation of the additional quantum noises and other undesirable negative effects with respect to the control laser fields and the predetermined atomic evolution.
Below we describe a basic model and main physical properties of the proposed QM scheme. Then we summarize the critical physical requirements and outline the possible experimental realizations.

\section{Basic model and equations}

We analyze interaction of three-level multi-atomic system with a single mode of QED cavity coupled with external signal field modes.
It is assumed that all atoms ($j = 1,2,...,N )$ stay initially
on the long-lived ground states $|1_a\rangle=\prod_{j=1}^{N} \ket{1}_{j}$ so the state of light and atoms is $|\Psi_{in}\rangle=|\psi_f\rangle_{in}|1_a\rangle$
before we launch any weak signal light field which can contain many temporally separated weak light pulses or even single photon fields. We analyze a sequence of single photon wave packets prepared in the quantum state
$\left| {\psi _{f} } \right\rangle _{in} = \prod\nolimits_{k = 1}^M
{\hat {\psi }_k^ + (t - \tau_k )} \left| 0 \right\rangle $, $\hat {\psi }_k^ +
(t - \tau_k ) = \int_{ 0 }^\infty {d\omega _k f_k (\omega _k )\exp \{ -
i\omega _k (t - \tau_k )\}\hat {b}^{+} (\omega _k )} $; $f_k$ is a wave function in the frequency space normalized for pure single photon state $\int_{0}^\infty {d\omega _k \vert f_k (\omega _k )\vert ^2 = 1} $, M is a number of temporal modes,
$\hat b^{+} $ and $\hat b $ are the arising and decreasing operators of the frequency field modes $[\hat {b} (\omega '),\hat
{b}^{+} (\omega )] =  \delta (\omega ' - \omega )$,
$\left| 0 \right\rangle = \left| 0 \right\rangle_m  \left| 0 \right\rangle_f $ where
$\left| 0 \right\rangle_m $ and $\left| 0 \right\rangle_f $ are the vacuum states of the cavity mode and of the free propagating field; k-th photon mode arrives in the circuit at time moment $\tau_k $, time delays between the nearest photons are assumed to be large enough $(\tau_k - \tau_{k - 1} ) \gg \delta t_k$, $\delta t_k\approx \delta\omega_k^{-1}$ is a temporal duration of the k-th field mode, with spectral width $\delta\omega_k^{-1}\leq\delta\omega_f^{-1}$, where $\delta\omega_f$ is an character spectral width.

\begin{figure}
\includegraphics[width=0.4\textwidth,height=0.25\textwidth]{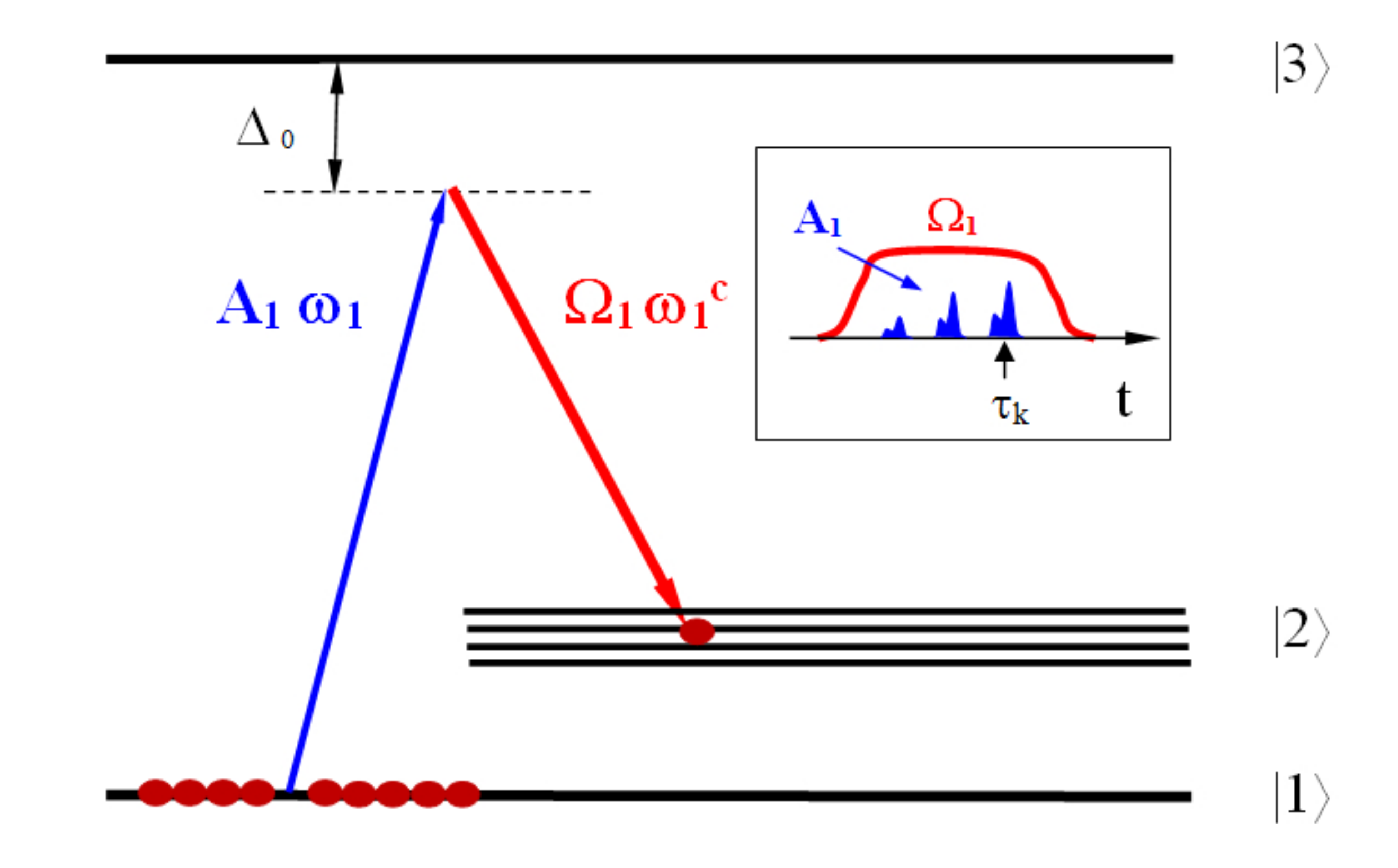}
\caption{Energy level diagram and nonresonant Raman transition on the atomic transitions
$\left| 1 \right\rangle \leftrightarrow\left| 2 \right\rangle $
due to the interaction with signal light pulse $A_{1}$ (with carrier frequency $\omega_{1}$) and with control (writing) laser field ($\Omega_{1}$ is a Rabi frequency on $\left| 2 \right\rangle \leftrightarrow\left| 3 \right\rangle $ transition and  $\omega_{1}^c $ is a carrier frequency of the control field), $\Delta_0$ is a sufficiently large resonant detuning from the optical transition. Inserted temporal diagram shows temporal shapes of the signal and control fields. }
\label{RammFig1}
\end{figure}

We assume that central carrier frequency of the input signal fields coincides with the QED cavity mode frequency $\omega_{f,0}=\omega_o$ and the atomic ensemble is simultaneously exposed to an intense control laser field characterizes by index $\nu=1$ propagating along wavevector $\vec{K}_{1}$ with carrier frequency $\omega_1^c$ and
Rabi frequency  $\tilde{\Omega} _{\nu} (t,\vec{r})=\Omega _{\nu} (t )$ $\exp \{ - i(\omega _{\nu}^c t-\vec {K}_{\nu} \vec {r})\}$
on $\left| 2 \right\rangle \leftrightarrow\left| 3 \right\rangle $ atomic transition
(where indexes $\nu=1,..,6$ are used for control laser fields in a storage ($\nu=1$), rephasing ($\nu=2,3,4,5$) and retrieval ($\nu=6$) stages of atomic and light evolution).
The excited cavity mode field $A_1$ and first control laser field $\Omega_1$ provides off-resonant Raman excitation of atoms with Lambda scheme of quantum transition as depicted in Fig.\ref{RammFig1}.

\begin{figure}
\includegraphics[width=0.4\textwidth,height=0.25\textwidth]{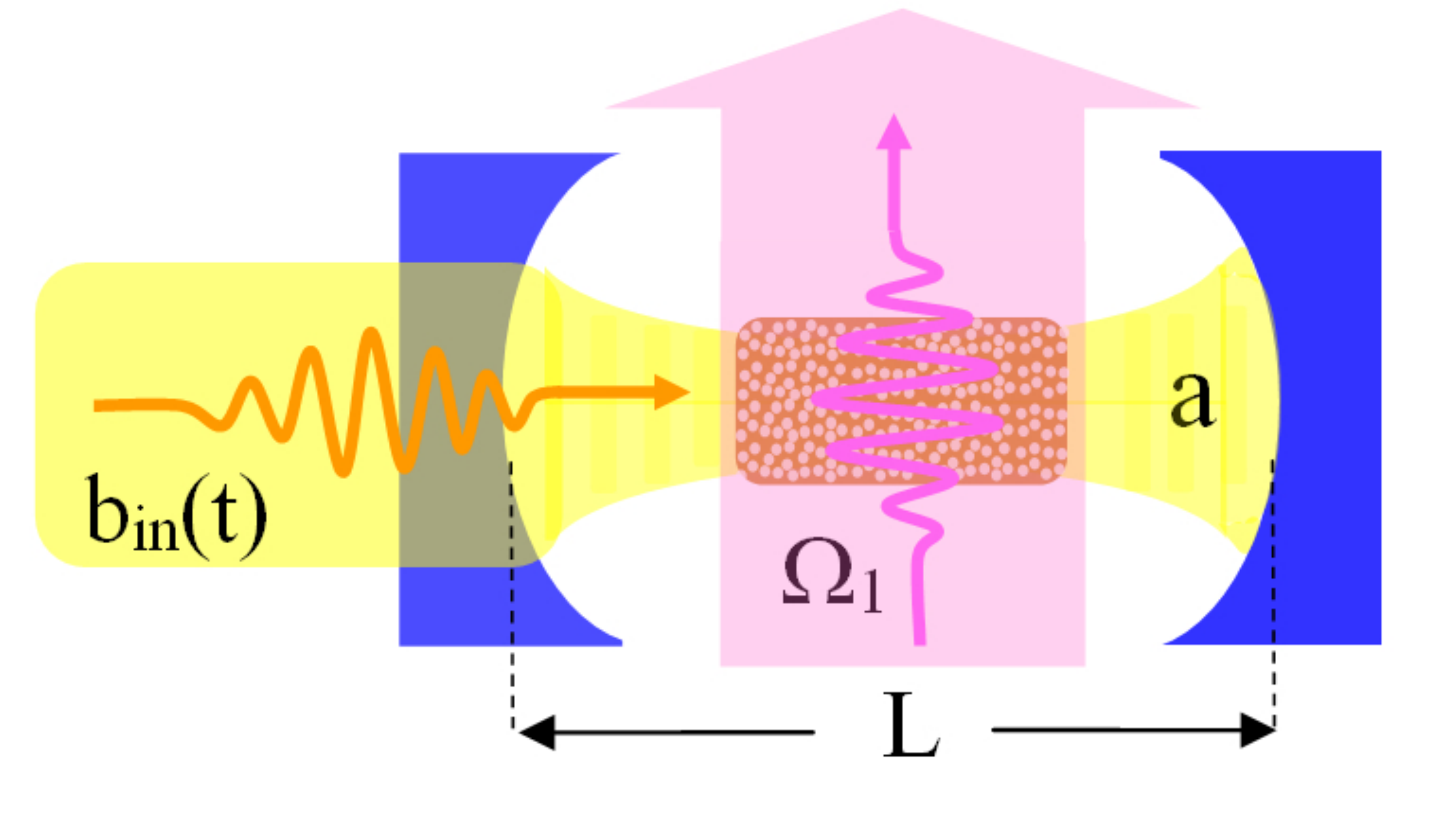}
\caption{Spatial diagram of interaction between the atomic ensemble and QED cavity mode $\hat a$ in a presence of the control classical laser field $\Omega_{1}$ propagating between the cavity mirrors, where the cavity mode is excited by input signal light field $\hat b _{in} (t)$.}
\label{RammFig2}
\end{figure}

Spatial diagram of the interaction presented in Fig.\ref{RammFig2} shows an excitation of atoms by the cavity mode field $\hat a$ and
control field $\Omega_1$ propagating through the atomic medium without any reflection from the cavity mirrors.
By following the cavity mode formalism for the coupling $\hat{V}_{m-f}$ of cavity mode with external light field modes  \cite{Milburn1994}, we use a Tavis-Cumming Hamiltonian \cite{Tavis1968} for N three-level atoms $\hat{H}_a$ and its interaction with cavity mode $\hat{V}_{a-m}$ and with control field $\hat{V}_{a-c}$.
Total atom-light Hamiltonian is:
\begin{eqnarray}
\label{Hamilt}
\hat{H}= \hat{H}_a + \hat{H}_m+\hat{H}_f + \hat{V}_{a-m}+\hat{V}_{a-c} +\hat{V}_{m-f},
\end{eqnarray}
where  $\hat{H}_a  = \hbar \sum^{N}_{j=1}[ \omega_{31} \hat{P}^{j}_{33}+ (\omega_{21}+ \delta^j)\hat{P}^{j}_{22}] \label{atomH} $ is a Hamiltonian of three-level atoms, $\hat{P}^{j}_{nm}=|n\rangle_{jj}\langle m|$  is a projection operator of $j$-th atom, $\delta^j$ is a spectral detuning on $\left| 1 \right\rangle \leftrightarrow\left| 2 \right\rangle $  transition of  $j$-th atom, $\hat{H}_m  = \hbar \omega_o   \hat{a}^{+}\hat{a}$  and  $\hat{H}_f=\hbar {\int {\omega\hat {b}^{+} (\omega )\hat{b} (\omega ) d\omega } }$ are the Hamiltonians  of the QED cavity mode and of free propagating modes coupled via the interaction term $\hat {V}_{m-f}$ and the interaction of atoms with  QED cavity mode and with classical control laser field are given by the
Hamiltonians  $\hat {V}_{a - m}$ and  $\hat {V}_{a-c}$:

\begin{eqnarray}
\label{Interac}
\hat {V}_{m-f}& = & - \hbar  {\int {d\omega [\kappa  (\omega ) \hat {b} (\omega )\hat {a}^{+} + h.c.]}},\nonumber\\
\hat {V}_{a-m}& = & - \hbar \sum\limits_{j = 1}^N \{g_j \hat {a} \hat {P}_{31}^j + h.c.\}, \nonumber\\
\hat {V}_{a-c}& = & - \hbar \sum\limits_{j = 1}^N
\{\Omega _{1} (t)\exp [ - i(\omega _{1}^c t - \vec{K}_{1} \vec{r}_j)] \hat {P}_{32}^j + h.c.\},
\end{eqnarray}

\noindent
where $\hat {a}^ + $ and $\hat {a}$ are arising and decreasing operators of the cavity mode,
$g_j=g_{j}^{o} \cos [\vec{k} \vec{r}_j]$, $g_{j}^{o}$ is the photon-atom coupling constant in QED cavity \cite{Scully1997},
$\vec{k}$ and $\vec{K_1}$ are the wave vectors of the standing cavity mode and of the $1$-st traveling control laser field,

By assuming a weak population of excited atomic states $\left| 2 \right\rangle$, $\left| 3 \right\rangle$ (i.e. $<\hat P_{22}^j>\ll1$ and
$<\hat P_{33}^J>\ll1$) due to the interaction of multi-atomic ensemble ($N\gg1$) with weak single photon fields we will use the input-output field formalism \cite{Milburn1994} and derive the linearized system of
Heisenberg equations for the field operators and for the atomic operators in the rotating frame representation
($\hat {b} (\omega )=\hat {b}_o (\omega )e^{-i\omega_s t} $, $\hat {a} (\omega )=\hat {a}_o (\omega )e^{-i\omega_s t}$,
$\hat {P}_{13}^j =\hat {P}_{o,13}^je^{-i\omega_s t}$, $\hat {P}_{12}^j =\hat {P}_{o,12}^je^{-i\omega_{21} t}$ where $\omega_s$ is a carrier frequency of the input signal field $\hat b(t)$) satisfying the condition of resonant Raman transition ($\omega_s-\omega _{1}^c \approx\omega_{21}$):

\begin{equation}
\label{b-eq}
\textstyle{d \over {dt}}\hat {b}_o (\omega ) = - i(\omega - \omega _s )\hat
{b}_o (\omega ) + i \kappa _l (\omega )\hat {a}_o,
\end{equation}

\begin{equation}
\label{a-eq}
\textstyle{d \over {dt}}\hat {a}_o = - i(\omega_o-\omega_s-i\textstyle{1 \over 2} \gamma _1) \hat {a}_o +
 i {\sum\nolimits_{j =1}^{N } {g_{j } } \hat {P}_{o,13}^j }  + \sqrt {\gamma _1 } \hat {b}_{in} (t),
\end{equation}

\begin{equation}
\label{13-eq}
\textstyle{d \over {dt}} \hat {P}_{o,13}^j   =  - i(\Delta _{31 }^j+\Delta_o) \hat {P}_{o,13}^j + i g_{j}^\ast \hat {a}_o ,
 + i \Omega_1(t) \exp [  i \vec{K}_{1} \vec{r}_j] \hat{P}_{o,12}^j,
\end{equation}

\begin{equation}
\label{12-eq}
\textstyle{d \over {dt}} \hat {P}_{o,12}^j  =  -i \delta^j \hat{P}_{o,12}^j
+ i \Omega_1 ^\ast(t) \exp [ - i \vec{K}_{1} \vec{r}_j]\hat{P}_{o,13}^j,
\end{equation}

\noindent
where $\gamma _l = 2\pi \kappa ^2 (\omega _o )$, the input signal field containing M temporally separated photon wave packets
is given by
$\hat {b}_{in} (t) = \sum\nolimits_{k = 1}^M {\hat {b}_{o,k} (t - \tau_k )}$ ,
where $\hat {b}_{o,k} (t - \tau_k) = \textstyle{1 \over {\sqrt {2\pi }}}
\int {d\omega \hat {b}_o (\omega ) \exp \{ - i(\omega - \omega _s )(t - \tau_k) \} }$ .

Let us assume that off resonant Raman interaction of the cavity mode with atoms occurs for large enough spectral detuning between the carrier frequency of signal field $\omega_{s}$ and the frequency $\omega_{31}$ of  optical atomic transition
$\left| 1 \right\rangle \leftrightarrow\left| 3 \right\rangle $ :
$\Delta_o =\omega_{31}-\omega_{s}$, $|\Delta_o|\gg\delta\omega_f$. It is well-known that in the case of sufficiently large spectral detuning $\Delta_o\gg\Delta_{in}^{(31)}$ (where $\Delta_{in}^{(31)}$ is IB of the transition
$\left| 1 \right\rangle \leftrightarrow\left| 3 \right\rangle $) the excited optical atomic coherence evolves adiabatically
following the cavity field and the long-lived atomic coherence in accordance with Eq. (\ref{13-eq}):

\begin{equation}
\label{adiabat}
\hat{P}_{o,13}^j (t) |\Psi\rangle_{in}\cong
\frac{1}{\Delta_{o}}\{g_{j}^\ast \hat {a}_o + \Omega_1(t) \exp [  i \vec{K}_{1} \vec{r}_j] \hat{P}_{o,12}^j \}|\Psi_{in}\rangle,
\end{equation}

\noindent
where we have taken into account the initial state $|\Psi_{in}\rangle$.

Total IB of the Raman transition will be determined only by inhomogeneous broadening $G (\delta/\delta_{in})$ on the atomic transition
$\left| 1 \right\rangle \leftrightarrow\left| 2 \right\rangle $ ($\delta_{in}$ is an IB linewidth of the atomic transition) where the transition can be forbidden that is highly preferable for many aspects of quantum storage (see below). It is worth noting
the adiabatic evolution (\ref{adiabat}) requires a large spectral detuning  $|\Delta_o|>|\Omega_1|$ and we can use  $|\Omega_1/\Delta_o|\approx 0.1$ in order to provide highly efficient and fast switching rate of the control field (principally we can use even larger spectral detuning  $|\Omega_1/\Delta_o|< 0.1$ by appropriate optimization of the atomic and light parameters).  Factor $|\Omega_1/\Delta_o|\ll 1$ decreases an effective optical depth of the atomic system, however this problem is resolved here due to considerable enhancement of light-atoms interaction in the QED cavity.

Putting Eq. (\ref{adiabat}) in Eqs. (\ref{a-eq}), (\ref{12-eq}) and assuming
$\omega_s=\tilde\omega_o$, $\delta^j=\Delta_{21}^j-\Omega_1/\Delta_o$, $\tilde g_{j }=g_{j } \exp [  i \vec{K}_{1} \vec{r}_j]$ and $\tilde\omega_o=\omega_o-N \overline{g}^2/\Delta_o$
(where $\overline{g}=\sqrt{\sum\nolimits_{j =1}^{N } {|\tilde g_{j }|^2}/N}$)
we get two light-atoms equations

\begin{equation}
\label{a-eq2}
\textstyle{d \over {dt}}\hat {a}_o = - \textstyle{1 \over 2} \gamma _1 \hat {a}_o +
 i \frac{\Omega_1(t)}{\Delta_o}{\sum\nolimits_{j =1}^{N } {\tilde g_{j }}  \hat {P}_{o,12}^j }
 + \sqrt {\gamma _1 } \hat {b}_{in} (t),
\end{equation}

\begin{equation}
\label{12-eq2}
\textstyle{d \over {dt}} \hat {P}_{o,12}^j  =  -i (\delta^j-i\frac{1}{T_2}) \hat{P}_{o,12}^j
+ i \frac{\Omega_1^\ast(t)}{\Delta_o} \tilde g_{j }^\ast \hat{a}_{o},
\end{equation}

\noindent
where for generalization we have also introduced a small decay constant $1/T_2$ of the atomic coherence
determined by the interaction of atoms with surrounding host atoms and with bath electromagnetic fields.
Coupled system of Eqs. (\ref{b-eq}),(\ref{a-eq2}),(\ref{12-eq2}) describe the interaction of QED cavity mode and resonant three-level atomic system under the action of input light signal $\hat b_{in} (t)$. The equations coincide formally with the light-atom equations for the photon echo QM on two-level atoms in QED cavity \cite{Moiseev2010} if we replace the photon-atom coupling constant $g_j$ by the controlled effective coupling constant $i\frac{\Omega_1^\ast(t)}{\Delta_o} \tilde g_{j}$.

\section{Quantum storage}

By assuming the control field is switched on to the constant magnitude $\Omega_1$ before the signal pulse arrives in the QED cavity at $t=0$,
we realize a storage of the input signal on the atomic coherences $\hat {P}_{o,12}^j$ and $\hat {P}_{o,13}^j$ after complete arrival of the
all M signal light pulses $\hat {a}_o(t>\tau_M+\delta t_M)|\Psi_{in}\rangle=0$. Further adiabatic switching of the control laser field for $t>\tau_M+\delta t_M+\tilde\tau $ leads to

\begin{equation}
\label{coher-st}
\hat {P}_{o,12}^j (t)= i \sqrt{2\pi}\frac{\Omega_1^\ast}{\Delta_o } \tilde g_{j}^\ast \hat a_o (\delta^j)
e^{-i( \delta^j-i/T_2) t},
\end{equation}

\begin{equation}
\label{a-st}
\hat a_o (\nu)= \frac{\sqrt{\gamma_1} \hat b_{in}(\nu)}{\frac{1}{2}\gamma_1-i\nu
-i\frac{1}{2} \Gamma_r \delta_{in}\int \frac{d\delta G(\delta/\delta_{in})}{\delta-\nu-i/T_{2}}},
\end{equation}

\noindent
where $\hat a_o (\nu)=\frac{1}{\sqrt{2\pi}}\int dt e^{i\nu t} \hat a_o (t)$,
$\hat b_{in} (\nu)=\frac{1}{\sqrt{2\pi}}\int dt e^{i\nu t} \hat b_{in} (t)=\sum_{k=1}^M e^{i\nu\tau_k}\hat b_{o} (\nu)$,
$\Gamma_r =2 N |\frac{\Omega_1 }{\Delta_o}\bar g|^2 / \delta_{in}$, and $\hat {P}_{o,13}^j (t)|\Psi_{in}\rangle=0$.

By using for simplicity the Lorentzian shape of IB $G_g(\delta/\delta_{in})=\frac{\delta_{in}}{\pi(\delta^2+\delta_{in}^2)}$ in Eqs. (\ref{coher-st}), (\ref{a-st}) we calculate a storage efficiency of the signal field
$Q_{ST}(t) = \bar {P}_{22}(t) / \bar {n}_1 $ where
$\bar {P}_{22}(t) = \sum\nolimits_{j = 1}^{N } { \langle \hat {P}_{o,21}^j (t) \hat {P}_{o,12}^j (t) \rangle } $
is an excited number of atoms after the interaction with  M
signal fields for $t > \tau_{M}+\delta t_M$
(where $\delta t_M \approx\delta\omega_f^{-1}$).
 Total number of photons
in the input signal field is $\bar {n} =
\sum\nolimits_{k = 1}^M {\bar {n}_{k} } $, where  $\bar {n}_{k} = \int_{ -
\infty }^\infty {dt \langle \hat {b}_{o,k}^ + (t)} \hat {b}_{o,k} (t)\rangle $ is the input number of photon in k-th temporal mode, $\langle...\rangle$ is a
quantum averaging over the initial state $\left| {\Psi_{in}}\right\rangle$.

Performing the algebraic calculations of $\bar {P}_{22}(t) $, we find the quantum
efficiency of storage $Q_{ST} = (1 / \bar {n}_1 )\sum\nolimits_{k = 1}^n {Q_{ST,k} }
\bar {n}_{k} $ where the storage efficiency  of k-th mode is

\begin{eqnarray}
\label{qe k_f}
Q_{ST,k} = &
\int\limits_{ - \infty }^\infty {d\nu }
{\rm Z}_r (\nu ,\delta _{in}, T_2,\gamma_1, \Gamma _{r} )
\frac{\left\langle {\hat {n}_{k} (\nu)}\right\rangle}{\bar {n}_{k}},
\end{eqnarray}

\noindent
where \emph{spectral storage }(SS-) function
${\rm Z}_r (\nu ,\delta _{in}, T_{2}, \gamma_1, \Gamma _{r} )=|S_r (\nu ,\delta _{in}, T_{2}, \gamma_1, \Gamma _{r} )|^2 $ and S-function $S_r (\nu ,...)$ is

\begin{eqnarray}
\label{Sp_f}
&{S}_r (\nu ,\delta _{in}, T_{2}, \gamma_1, \Gamma _{r} ) =
\sqrt{\frac{\delta _{in}^2 }{(\delta_{in}^2 + \nu ^2)}}
\frac{2\sqrt{\gamma _1 \Gamma _{r}} }{[ \gamma _1 + \Gamma _{r}\frac{\delta _{in}}{(\delta _{in}+1/T_2 - i\nu)}- 2i\nu ]},
\end{eqnarray}

\noindent
characterizes the quantum storage $Q_{ST,k}$ for $\nu$-th spectral component of $k$ -th light temporal mode.
It is worth noting SS-function ${\rm Z}_r (\nu,...) $ differs only by new parameter $\Gamma_r$ from the similar
SS-function in the photon echo QM
on two-level atoms studied in \cite{Moiseev2010}, where $\Gamma_r$
characterizes a photon absorption rate on the off resonant Raman transition. So by assuming $1/T_2\ll\delta_{in}$ from Eqs. (\ref{qe k_f}), (\ref{Sp_f}) we get the following two conditions for realization of efficient quantum storage:

1) $\Gamma_r =\gamma_1$ and 2) $\delta_{in}=\gamma_1/2$.

The first one 1)  is an well-known impedance matching condition in laser physics \cite{Haus1984, Siegman1986} modified by the Raman atomic transition. By taking into account $\gamma_1 \approx T c/(2L)$ (where $T=1-R$ and $R$ are the transmission and refraction coefficients (for the left mirror in Fig.2 ),  $c$ is a speed of light, $L$ is a longitudinal  length of the QED cavity), we get for the transmission coefficient $T=2 \chi  \alpha_{r} L$, where the absorption coefficient on the Raman transition
$\alpha_{r} = |\frac{\Omega_1 }{\Delta_o}|^2 \frac{\Delta_{in}^{(13)}}{\delta_{in}}\alpha_{13}$ and
$\alpha_{13} = 4 \pi n_o  \frac{|d_{13}|^2\omega_{31}}{c \hbar \Delta_{in}^{(13)}}$ is an absorption coefficient on the optical transition,
$n_o=\frac{N}{V_a}$ - atomic density, $V_a$ - volume of atomic medium,  $\chi=V_{a}/(LS)$ is a filling factor of the cavity
by the atomic system, $S$ is an averaged cross-section of the cavity mode.

Thus instead of the strick demand on the resonant optical depth $\alpha_r L \gg 1$ in the case of free space QM \cite{Moiseev2001}, we have a quite weaker demand even for the Raman scheme of atomic transition.  We rewrite the condition for absorption coefficient  $\alpha_r = T/(2\chi L)$ which can be easily realized in experiments with optical QED cavities.
To give an example let us assume $\gamma_1=10^8 sec^{-1}$ and $L=0,1$ cm (or $1$ cm), we get $T \cong 0.7 \cdot 10^{-3}$ (or $T \cong 0.7 \cdot 10^{-2}$).
By assuming also $\chi=0.5$  and $|\frac{\Omega_1 }{\Delta_o}|^2 = 0.01$ we get the matching condition for the optical absorption coefficient $\alpha_{13}=0.7 \cdot \delta_{in}/\Delta_{in}^{(13)}$ independent of $L$.
Usually $\delta_{in}/\Delta_{in}^{(13)}$  lays in the range $0.1 \div 0.01$ for rare-earth ions in the inorganic crystals \cite{Tittel2010} where $\Delta_{in}^{(13)} \approx 10^9\div 10^{10}$ sec$^{-1}$.
Thus the matching condition clearly indicates a possiblilty of using the atomic systems with a quite small optical absorption coefficients $\alpha_{13}<0.1$ and even $\alpha_{13}<0.01$ that means a lower density of the resonant atoms and weaker interatomic dipole-dipole interactions, respectively, that also promises a larger lifetime of the atomic coherence $T_2$. We also note that we can vary IB linewidth $\delta_{in}$ in external constant electrical or magnetic fields without using CRIB procedure for perfect retrieval.

It is worth noting off resonant Raman interaction facilitates an experimental realization of the impedance matching condition due to the possible tuning of Rabi frequency  $\Omega_1$ and spectral detuning $\Delta_o$: $|\frac{\Omega_1}{\Delta_o}|=\sqrt{\frac{\delta_{in}\gamma_1}{2N|\bar g|^2}}$
which is important for the atomic media with arbitrary number N of effective three-level atoms.

Another well-known important advantage of the Raman scheme is a direct mapping of the input light fields on the long-lived atomic transition
$\left| 1 \right\rangle\leftrightarrow \left| 2 \right\rangle $ so the transition can be completely forbidden for realization of spontaneous coherent atomic irradiation on this transition and this atomic transition will suppress many relaxation processes providing a large $T_2$ for  long-lived storage of the input signal fields.

The second condition 2) called \emph{optimal spectral matching condition} \cite{Moiseev2010} demonstrates an optimal coupling of the spectral shapes related to the atomic IB and  QED cavity window transparency which provides an effective quantum storage for broadened spectral range of input signal fields where the SS-function ${\rm Z}_r (\nu,...)\cong 1 $.

The excited atomic systems is characterized by completely dephased coherence (\ref{coher-st}) on the transition
$\left| 1 \right\rangle\leftrightarrow \left| 2 \right\rangle $
and the light field retrieval requires a perfect rephasing of the atomic coherence for subsequent irradiation of the echo field.  Let us now consider the retrieval stage.

\section{Retrieval stage}

\subsection{Rephasing of atomic coherence}

Retrieval still remains the most critical stage in the photon echo QMs. Here we propose an efficient solution of this problem by using the specific advantages of Raman based techniques for the atomic coherent control in optical QED-cavity.
Moreover we implement the retrieval for the atomic system with natural IB of atomic transition $\left| 1 \right\rangle \leftrightarrow\left| 2 \right\rangle $ without any dynamical or static manipulations of the atomic spectral detunings used in CRIB \cite{Moiseev2001} and in AFC \cite{Riedmatten2008} techniques for rephasing of the excited IB atomic coherence. Despite the recently proposed new photon echo QM techniques also based  on the natural IB \cite{Moiseev2011, McAuslan2011,Damon2011,BHam2012}, the efficient experimental realization of these schemes still requires additional improvements which would eliminate the specific physical shortcomings caused by extra quantum noises and limited quantum efficiency.

\begin{figure}
\includegraphics[width=0.3\textwidth,height=0.3\textwidth]{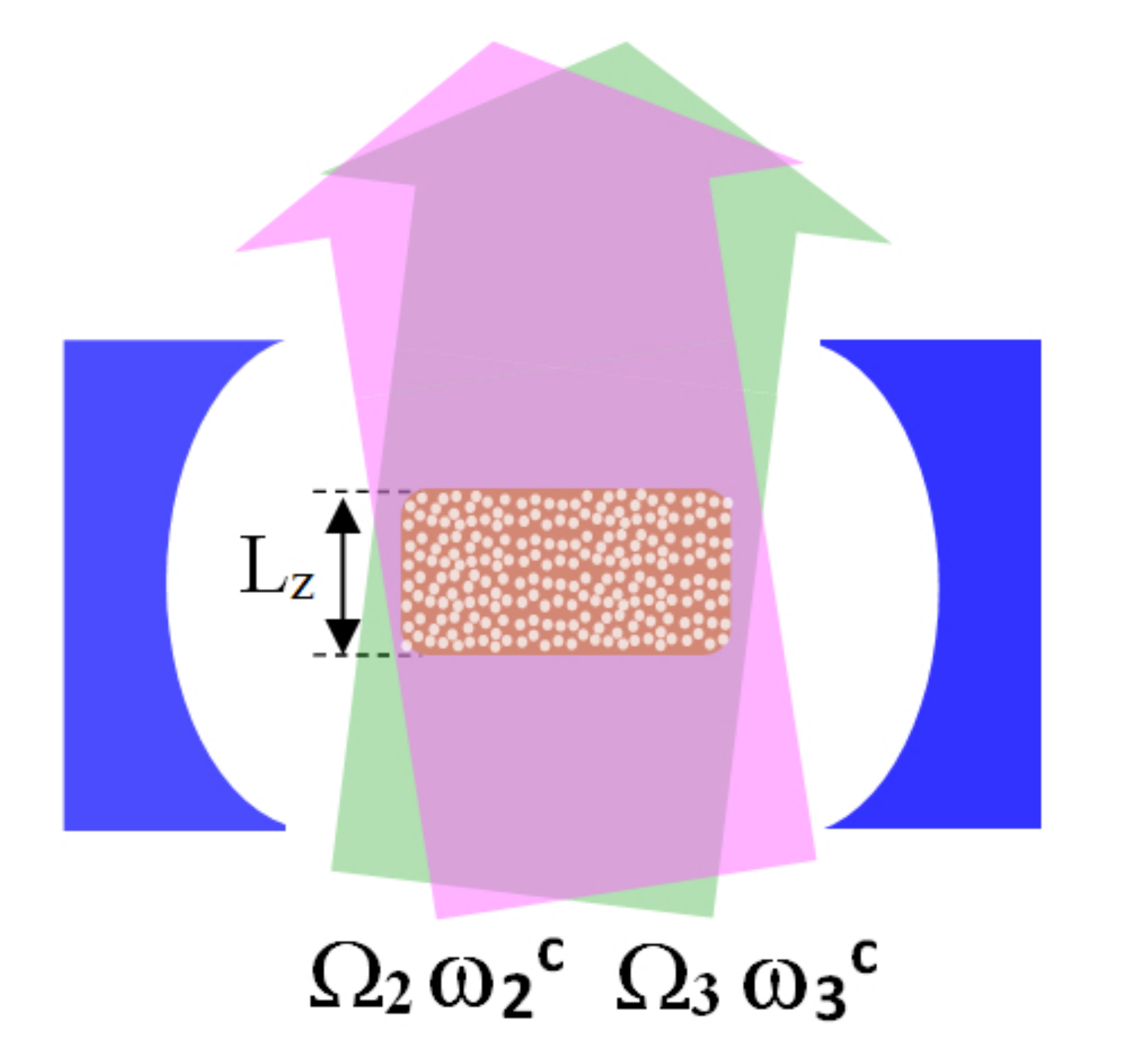}
\caption{Spatial scheme of the atomic excitation by two control fields $\Omega_{2}$ and $\Omega_{3}$  satisfying the Raman resonance with atomic transition $\left| 1 \right\rangle \leftrightarrow\left| 2 \right\rangle $: $\omega_{2}^c-\omega_{3}^c\approx\omega_{21}$. The fields can propagate in arbitrary directions between the cavity mirrors.}
\label{RammFig3}
\end{figure}

\begin{figure}
\includegraphics[width=0.4\textwidth,height=0.3\textwidth]{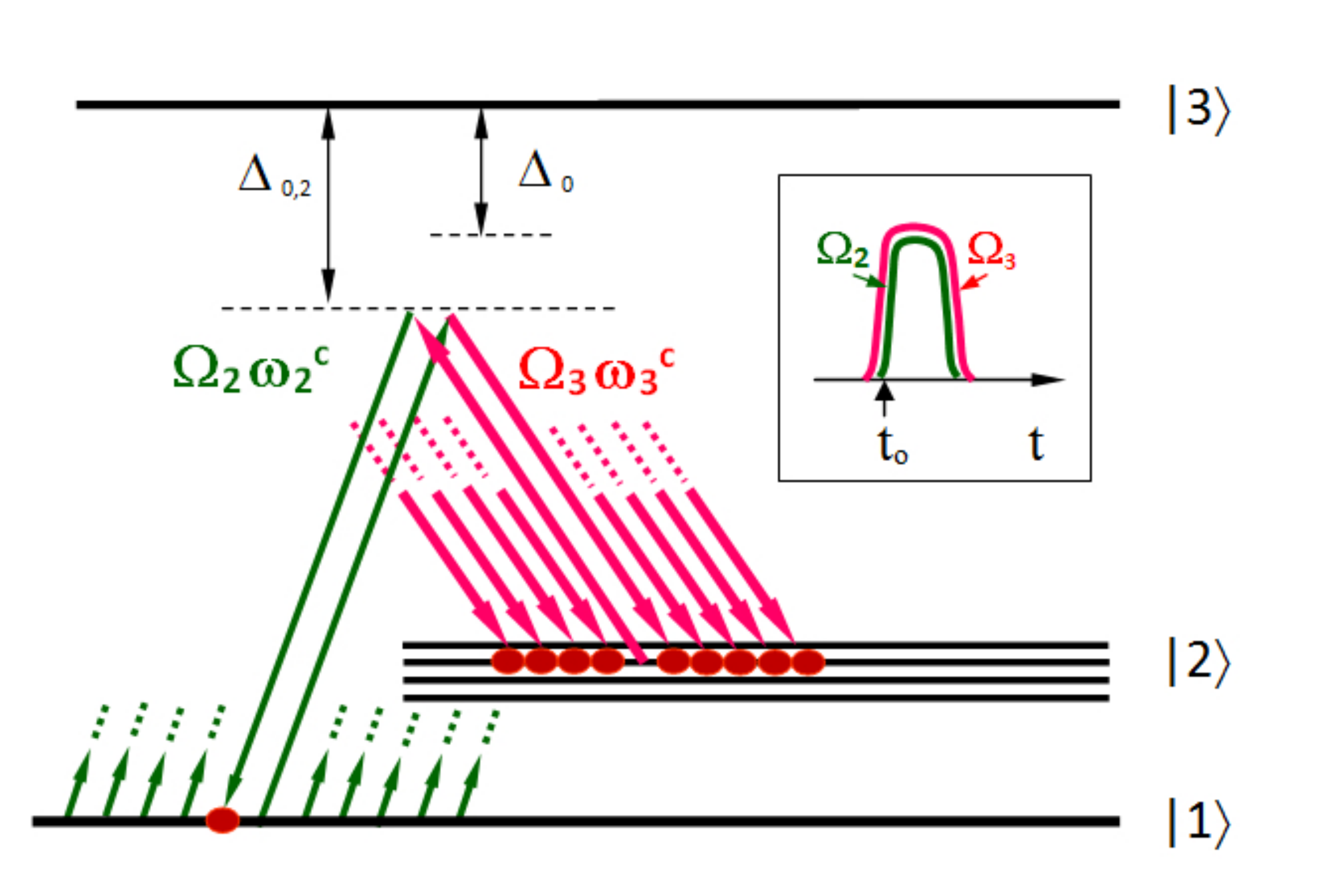}
\caption{Two control lasers $\Omega_{2}$ and $\Omega_{3}$ invert atomic state on the transition
$\left| 1 \right\rangle \leftrightarrow\left| 2 \right\rangle $
via  off-resonant Raman transition  characterized  by another spectral detuning on the optical transition ($\Delta_{0,2} \neq \Delta_0$ so the laser field frequencies do not coincide with the QED cavity frequency: $\omega_{2}^c \neq \omega_{o}$ $\omega_{3}^c\neq \omega_{o}$); the arrows demonstrate perfect $\pi$-pulse area of the excitation for the atoms. The insert figure shows temporal diagram for two control rephasing laser fields ($\nu=2,3$) with coincided temporal shapes.}
\label{RammFig4}
\end{figure}

Our scheme makes use of the additional experimental tools for realization of perfect manipulation of the atomic coherence providing highly efficient rephasing of the long-lived coherence $\hat P_{o,12} (t)$ (\ref{coher-st}) excited on the natural IB transition.
To start the coherence rephasing at time $t=t_o$ we launch two intensive laser fields ($\nu=2,3$) exciting off resonant Raman atomic transition
$\left| 1 \right\rangle\leftrightarrow\left| 2 \right\rangle$ in the following way.
As is shown in the spatial scheme in Fig. \ref{RammFig3}, the laser pulses simultaneously propagate between the two mirrors of QED cavity. It is taken into account the pulses can propagate in arbitrary angle to each others (they could be even in opposite directions). This can be realized for the cases discussed in the previous section, where for example the distance between two mirrors of the QED cavity can range  from $1$ mm. to $1$ cm. We also assume the input laser pulses parameters provide a perfect quantum transition $\left| 1 \right\rangle \leftrightarrow\left| 2 \right\rangle $ characterized by the pulse area equal to $\pi$. Also we can implement the excited transient optical coherences $\hat P_{o,13} (t)$ and $\hat P_{o,23} (t)$ are not coupled with any QED-cavity modes since the carrier frequencies of the laser pulses $\omega_2^c$ and $\omega_3^c$ can be shifted from the resonant frequencies of the QED-cavity modes as depicted in Fig.\ref{RammFig4}.

In comparison with the properties of usual photon echo QM schemes, we show the control intensive light fields will propagate through medium in our scheme without significant backward influence of the atomic medium, i.e. the atomic system can be characterized by negligibly small optical depth for the control laser fields.
In order to clarify the noted key scheme advantages we examine the light-atoms dynamics for the case of short control light pulses and then we characterize the optimal atomic parameters. Similarly to the storage stage we assume a sufficiently large optical detuning $\Delta_{0,2}=\omega_{31}-\omega_2^c$ (see Fig. \ref{RammFig4} where $|\Delta_{0,2}|>\Delta_{0}$) that determine adiabatic relations for the optical coherences

\begin{eqnarray}
\hat {P}_{o,13}^j\cong\frac {1}{\Delta_{0,2}} \{\Omega_2 (t) e^{ik_2 r_j}(\hat{P}_{o,11}^j-\hat{P}_{o,33}^j)
+ \Omega_3 (t) e^{ik_3 r_j}\hat{P}_{o,12}^j \}, \nonumber \\
\hat {P}_{o,23}^j\cong\frac {1}{\Delta_{0,2}} \{\Omega_3 (t)e^{ik_3 r_j} (\hat{P}_{o,22}^j-\hat{P}_{o,33}^j)
+ \Omega_2 (t) e^{ik_2 r_j}\hat{P}_{o,21}^j \},
\end{eqnarray}

\noindent
and leads to the dynamic equations for atomic evolution determined by the well-known effective Hamiltonian $\hat H_{eff,1} (t)=\sum_{j=1}^N \hat H_{eff,1}^j(t)$ where $\hat H_{eff,1}^j(t)=\hbar \delta^j(t)\hat P_{22}^j+\hat V_j(t)$ and the light-atom interaction term

\begin{equation}
\label{Eff_Hamil}
\hat V_j(t)=- \hbar\{\frac{\Omega_{2,j}(t)\Omega_{3,j}^*(t)}{\Delta_{0,2}} \hat P_{0,12}^j e^{-i(k_2-k_3)r_j}+h.c. \},
\end{equation}

\noindent
where $\delta^j(t)=\delta^j -\frac{|\Omega_{2,j}(t)|^2-|\Omega_{3,j}(t)|^2}{\Delta_{0,2}}$ $\Omega_{2,j}(t)=\wp_{13} E_2 (t,r_j)/\hbar$ is an atomic detuning, $\Omega_{3,j}(t)=\wp_{23} E_e (t,r_j)/\hbar$ are Rabi frequencies for $j$-th atom of the second and third control laser pulse on $\left| 1 \right\rangle \leftrightarrow\left| 3 \right\rangle $ and
$\left| 2 \right\rangle \leftrightarrow\left| 3 \right\rangle $ atomic transitions, $\wp_{13}$ and $\wp_{23}$ are the appropriate dipole moments of the atomic transitions.

Effective Hamiltonian $\hat H_{eff,1}^j$ determines the following atomic equations

\begin{equation}
\label{12_at_reph}
\textstyle{d \over {dt}} \hat {P}_{o,12}^j  =  -i (\delta^j-\frac{|\Omega_2(t)|^2-|\Omega_3(t)|^2}{\Delta_{0,2}}) \hat{P}_{o,12}^j
+ i \frac{\Omega_2 (t) \Omega_3^*(t)}{\Delta_{0,2}} e^{i(k_2-k_3)r_j}\hat W^j,
\end{equation}

\begin{equation}
\label{22_at_reph}
\textstyle{d \over {dt}} \hat {W}^j  =
-2 i (\frac{\Omega_2 (t) \Omega_3^* (t)}{\Delta_{0,2}}\hat{P}_{o,21}^j e^{i(k_2-k_3)r_j}-h.c.),
\end{equation}

\noindent
where $\hat W^j = \hat{P}_{o,11}^j-\hat{P}_{o,22}^j$,  $\textstyle{d \over {dt}} \hat {P}_{o,11}^j=-\textstyle{d \over {dt}} \hat {P}_{o,22}^j$,
equations for $\hat {P}_{o,nm}^j$  ($n,m=1,2,3; n\neq m$) are the hermitian conjugated equations for $\hat {P}_{o,mn}^j$.

By assuming a negligibly small angle between the propagation directions we write the equations for slowly varied amplitudes of the control light fields $E_2(t,z)$ and $E_3(t,z)$:

\begin{equation}
\label{Cont F eq2}
(\textstyle{\partial \over {\partial z}} +\frac{1}{c} \textstyle{ \partial \over {\partial t} })E_2 (t,z)
=  i \beta_1 e^{-ik_2 z} <P_{0,13}(t,z)>  ,
\end{equation}

\begin{equation}
\label{Cont F eq3}
(\textstyle{\partial \over {\partial z}} +\frac{1}{c} \textstyle{ \partial \over {\partial t} })E_3 (t,z)
=  i \beta_2 e^{-ik_3 z} <P_{0,23}(t,z)>  ,
\end{equation}

\noindent
where $\beta_{p}=\frac{2\pi}{c}  \omega_{3p} n_o \omega_{3p} d_{p3} $, ($p=1,2$), $<P_{0,13}(t,z)>$ and $<P_{0,23}(t,z)>$ are the macroscopically averaged atomic coherences, $z$-axis is orthogonal to the longitudinal axis of the QED-cavity (see Fig. 3).

Let us analyze the propagation effects of the fields $E_2 (t,z)$ and $E_3 (t,z)$ by assuming a short input temporal durations of the laser pulses in order to neglect by any influence  of IB on the transition $\left| 1 \right\rangle \leftrightarrow\left| 2 \right\rangle $ (i.e. $\delta_{in} \cdot\delta t_{2,3}\ll1$). Also we assume initial Rabi frequencies are equaled to each others ($\Omega_2(t)=\Omega_2^*(t)=\Omega_3(t)=\Omega_3^*(t)$). It leads to the following solution of Eqs. (\ref{12_at_reph}), (\ref{22_at_reph}):

\begin{equation}
\label{12_cont sol}
\hat {P}_{o,12}^j(t)= \cos^2{[\Theta (t)/2]}\hat {P}_{o,12}^j (\tau_o)+\sin^2{[\Theta (t)/2]}\hat {P}_{o,21}^j (\tau_o) e^{2i(k_2-k_3)r_j}
+i \frac{1}{2} e^{i(k_2-k_3)r_j} \sin{[\Theta (t)/2]} \hat {W}^j (\tau_o),
\end{equation}

\begin{equation}
\label{22_cont sol}
\hat {W}^j (t) = \cos{\Theta (t)}\hat {W}^j (\tau_o) +\sin{\Theta (t)} \{\hat{P}_{o,21}^j (\tau_o) e^{i(k_2-k_3)r_j}-\hat{P}_{o,12}^j(\tau_o) e^{-i(k_2-k_3)r_j}\},
\end{equation}

\noindent
where $\Theta (t)=2\int_{t_o}^{t} dt \frac{\Omega_2(t) \Omega_3(t)}{\Delta_{0,2}} $, $\hat {P}_{o,12}^j (\tau_o)$  is determined by initially excited atomic coherence (\ref{coher-st}), ($\hat {P}_{o,21}^j (\tau_o)=(\hat {P}_{o,21}^j (\tau_o))^+$) and $\hat {P}_{o,22}^j (\tau_o)$ is determined appropriately.

By taking into account the solutions (\ref{12_cont sol}), (\ref{22_cont sol})  in Eqs. (\ref{Cont F eq2}),(\ref{Cont F eq3})  and negligibly weak initial population of excited levels ($<\hat {P}_{o,22}(\tau_o)>\ll1$, $<\hat {P}_{o,33}(\tau_o)>=0$)  due to excitation by the input signal light we write the field equations in the moving system of coordinate ($Z=z$, $\tau=t-z/c$) in terms of its Rabi frequencies

\begin{equation}
\label{Cont F eq2-2}
\textstyle{\partial \over {\partial Z}} \Omega_2
=  - \frac{1}{2}\alpha_{r,13}\{ \frac{\Omega_3}{2} sin{\Theta}-i\Omega_2 cos^2[\Theta/2] \} ,
\end{equation}

\begin{equation}
\label{Cont F eq2-3}
\textstyle{\partial \over {\partial Z}} \Omega_3
=  \frac{1}{2}\alpha_{r,23} \{ \frac{\Omega_2}{2} sin{\Theta}+i\Omega_3 sin^2[\Theta/2] \}.
\end{equation}

The nonlinear system of the control field equations (\ref{Cont F eq2-2}),(\ref{Cont F eq2-3}) show that possible influence of absorption(amplification) and dispersion effects are determined by the coupling constants $\alpha_{r,p3}=\frac{\Delta_{in}^{p3}}{\Delta_{0,2}}\alpha_{p3}$  (where $p=1,2$) and spatial length $L_z \cong S^{1/2}$ of the medium in the cross-section of QED cavity.
For evaluation we assume  equal IB widths $\Delta_{in}^{(23)}=\Delta_{in}^{(13)}$  and absorption coefficients $\alpha_{23}=\alpha_{13}$ in Eqs. (\ref{Cont F eq2-2}),(\ref{Cont F eq2-3}) which leads to the equation for effective pulse area $\Theta (\tau,Z)$

\begin{equation}
\label{Theta}
\textstyle{\partial^2 \over {\partial Z \partial \tau}} \Theta
= \frac{1}{2}\alpha_{r,13}
\{ \frac{\Omega_2^2-\Omega_3^2}{\Delta_{0,2}} sin{\Theta}+i\textstyle{\partial \over {\partial \tau}} \Theta \}.
\end{equation}

\noindent
By taking into account the equal control field magnitudes ($\Omega_2^2-\Omega_3^2=0$) we find the main influence of light-atom interaction leads to the phase rotation  $\Theta (\tau,Z)\cong\Theta (\tau,0)\exp\{i\frac{1}{2}\alpha_{r,13}Z \}$ that can change the phase-matching condition in the medium only in case of sufficiently large effective optical depth $\frac{1}{2}\alpha_{r,13}L_z$. Similar evolution will occur for other parameters of the control fields, such as $\Omega_2(\tau,Z)$, $\Omega_3(\tau,Z)$ and $\Omega_2^2(\tau,Z)-\Omega_3^2(\tau,Z)$.

We can estimate the evolution of control light field parameters  by taking into account the optimal absorption coefficient and spatial properties of the atomic medium discussed in the previous section.
In accordance with the discussion, the first matching condition is determined by the longitudinal length of the cavity and absorption coefficient $\alpha_{13}$ so we can use a small cross-section of the QED-cavity with $L_z\ll L$ where minimal spatial length $L_z$ can be limited  by the a few wavelengths of the cavity mode. For experimental convenience we can use larger spatial length $L_z\approx L=1$ mm. Even for this case we have $\frac{1}{2}\alpha_{r,13}L_z\ll 1$  since $\frac{\Delta_{in}^{p3}}{\Delta_{0,2}}\ll 1$ and $\alpha_{r,13}L_z\ll1$ so with high accuracy the effective pulse area $\Theta$ will not be changed in the atomic medium. Moreover one can realize $\Theta (Z)=\Theta (0)$  with a priori given arbitrary accuracy since the negligibly small effective optical depth of the atomic system along $z-$direction can be realized by sufficiently high transverse confinement of the QED cavity mode field and appropriate transverse medium size. It is clear that in this case of effectively small optical depth we obtain the same result for different values of the absorption coefficients $\alpha_{r,13}$ and $\alpha_{r,23}$ and its IB widths. Thus we can neglect by any backward action of the atomic medium to the control light fields which provides a uniform evolution of the atomic coherences determined by the input parameters of control laser pulses with fixed equal real Rabi frequencies $\Omega_2(t)=\Omega_3(t)$ with an effective pulse area $\Theta$. Quantum dynamics of any $j$-th atomic operator in the laboratory system of coordinate will be determined as follows
$\hat P_{0,nm}^j (t=\delta t_2+t_o) = \hat U_j^+ (\Theta) \hat P_{0,nm}^j (t_o)\hat U_j (\Theta)$ where
$\hat U_j (\Theta)=\exp\{- i \frac{\Theta}{2}[\hat P_{0,12}^j e^{-i(k_2-k_3)r_j}+h.c.] \}$, $\delta t_2$ is a temporal duration of the control laser pulses (the temporal durations of the control laser pulses are assumed to be a negligibly short).

Here, we assume $\Theta=\pi$ that leads to the perfect inversion of atomic states on the transition $\left| 1 \right\rangle \leftrightarrow\left| 2 \right\rangle $. We note it is difficult realize such perfect inversion of all two-level atomic ensemble via applying intensive resonant light pulse in a single mode QED cavity due to many reasons. For example such $\pi$ pulse immediately leads to the effect of Dicke superradiance \cite{Dicke1954} and enhance strong interatomic interactions. In the free propagating scheme it leads to strong nonlinear light propagation effects of self-induced transparency in the echo irradiation \cite{Allen1975, Moiseev1987,Moiseev2004b} which can be accompanied by strong dispersion effects \cite{Damon2011} and even by soliton pulse formation \cite{Lamb1980, Rupasov1982}. Here it is also worth noting the atoms in the studied scheme are excited to the forbidden transition $\left| 1 \right\rangle \leftrightarrow\left| 2 \right\rangle $ that is also out of any resonance with QED cavity modes. Thus our scheme provides a robust perfect inversion of atomic states without essential inter-particle interactions and interaction with the QED cavity modes.

\begin{figure}
\includegraphics[width=0.65\textwidth,height=0.2\textwidth]{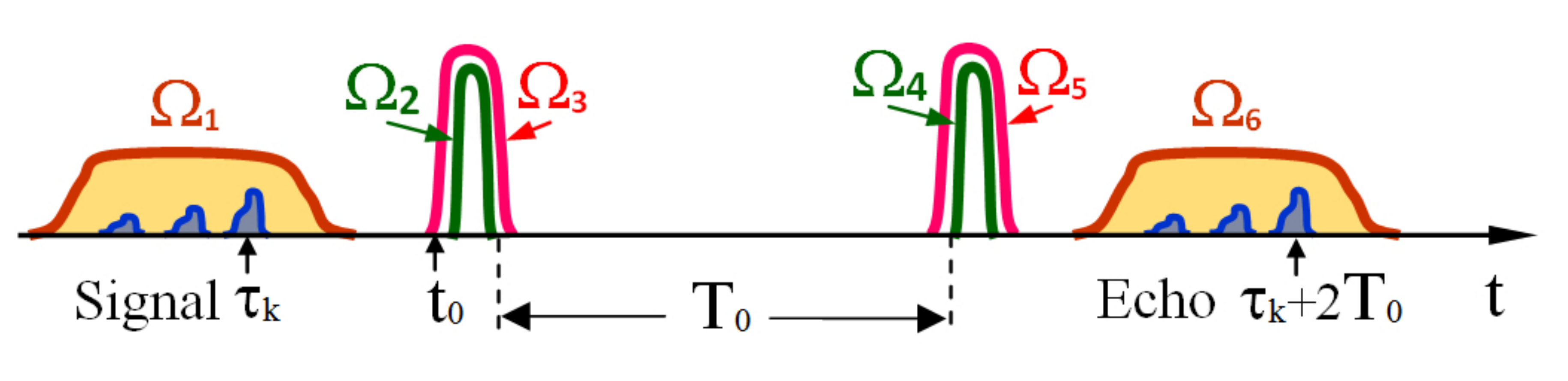}
\caption{Temporal scheme. The input signal (three blue (dark) pulses) enters the medium in the presence of $1$-st writing laser field
$\Omega_{1}$; two pairs of the control laser pulses $\Omega_{2},\Omega_{3}$ and $\Omega_{4},\Omega_{5}$ with effective pulse areas $\pi$ are applied for rephasing of the excited atomic coherence; reading laser pulse $\Omega_{6}$ initiates echo field emission characterized by the same temporal shape and order as it is for the input signal field.}
\label{RammFig5}
\end{figure}

Subsequent evolution of the atomic coherence will be determined by the Hamiltonian $\hbar \delta^j\hat P_{22}^j$ containing only different spectral atomic detuning
$\hat P_{0,12}^j (T_o+\delta t_2+t_o) =
\hat T_j^+ (T_o)  \hat P_{0,12}^j (\delta t_2+t_o)\hat T_j(T_o)$.
Complete procedure of the atomic rephasing contains action of two pairs of the control laser pulses separated by time delay $T_o>t_o-\tau_k$ (where $t_o-\tau_k$ is a dephasing time for $k$-th temporal mode (see Fig.\ref{RammFig5}) where each pair of laser pairs provides effective pulse area $\pi$ of the atomic excitation that leads to the following behavior of the long-lived atomic coherence

\begin{eqnarray}
\label{retr_coh}
\hat P_{0,12}^j (\delta t_4+T_o+\delta t_2+t_o)
= \hat U_j^+ (\pi) \hat T_j^+ [T_o]\hat U_j^+ (\pi)  \hat P_{0,12}^j (t_o)
\hat U_j (\pi)\hat T_j[T_o] \hat U_j (\pi)
 =  - e^{i\delta^j T_o}\hat P_{0,12}^j (t_o).
\end{eqnarray}

\noindent
Thus the recovered atomic coherence  $\hat P_{0,12}^j (t)$ at time moment $t=\delta t_4+T_o+\delta t_2+t_o$  acquires an additional phase $\delta^j T_o$ of the opposite sign with respect to the atomic phase excited after the signal field absorption. The rephasing process is highly immune to main negative physical processes usually limiting the quantum efficiency caused by light field propagation in the optical dense media and excitation of atoms on the excited allowed quantum states. Basically the accuracy of the rephasing is limited only by the accuracy of tuning of effective pulse area related to the control laser fields. In order to increase the accuracy of atomic inversion on the transition  $\left| 1 \right\rangle \leftrightarrow\left| 2 \right\rangle $ we can also apply  so-called counterintuitive pulse sequence in the well-known STIRAP scheme \cite{Scully1997} which could provide a robust perfect Raman transfer of atoms for relatively arbitrary intensities of the laser pulses.

\subsection{Echo field emission}

In order to retrieve the stored quantum state we launch an additional $6$-th reading laser pulse at time
$t >  t_o+ T_o$ (see Fig. \ref{RammFig5}) when the initial state of atomic coherence  is defined by  Eqs. (\ref{coher-st}) and (\ref{retr_coh}). The reading pulse switches on the coherent light-atoms interaction  due to the evolving rephasing of atomic coherence on the transition $\left| 1 \right\rangle \leftrightarrow\left| 2 \right\rangle $ when QED-cavity mode evolves from the vacuum state. Let us assume the reading laser pulse is characterized by Rabi frequency $\Omega_6$ and its carrier frequency coincides with the frequency of first (writing) control laser field ($\Omega_6(\tilde t)=\Omega_1(\tilde t)$, $\omega_6^c=\omega_1^c$). In this case after complete switching on, we obtain the same equations for the free propagating modes (\ref{b-eq}) and a similar system of equations for the atomic coherence $\hat P_{0,12}^j$ and QED-cavity mode Eqs. (\ref{a-eq2}), (\ref{12-eq2}) but without external light field source

\begin{equation}
\label{echo}
\textstyle{d \over {dt}}\hat {a}_o = - \textstyle{1 \over 2} \gamma _1 \hat {a}_o +
 i \frac{\Omega_6}{\Delta_o}{\sum\nolimits_{j =1}^{N } {\tilde g_{j }}  \hat {P}_{o,12}^j },
\end{equation}

\begin{equation}
\label{12-reph}
\textstyle{d \over {dt}} \hat {P}_{o,12}^j  =  -i (\delta^j-i\frac{1}{T_2}) \hat{P}_{o,12}^j
+ i \frac{\Omega_6^\ast}{\Delta_o} \tilde g_{j }^\ast \hat{a}_{o},
\end{equation}

\noindent
By using the Laplace transformations in Eqs.(\ref{echo}),(\ref{12-reph}) we find the solution for the QED-cavity mode field
$\hat {a}_o (t)=\sum_{k=1}^{M} \exp\{-(t-\tau_k)/T_2\}\hat {a}_{e,k} (t) $ where

\begin{eqnarray}
\label{echo-sol}
&\hat {a}_{e,k} (t) =
\frac{1}{\sqrt{2\pi \gamma_1}}
\int_{-\infty}^{\infty}d\nu \hat b_o (\nu)
\nonumber \\ &
\times
[{S}_r (\nu ,\delta _{in}, T_{2}, \gamma_1, \Gamma _{r} )]^2
\exp\{-i\nu(t-2 T_0 -\tau_k) \}
\end{eqnarray}

\noindent
Then using (\ref{echo-sol}) in (\ref{b-eq}) we find for the echo field
$\hat b_o (t) = \sqrt{\gamma_1} e^{-2T_o/T_2} \sum_{k=1}^{M}\hat {a}_{e,k} (t)$.
Thus we see if S-function ${S}_r (\nu ,\delta _{in}, T_{2}, \gamma_1, \Gamma _{r} )=1$, the echo field $\hat b_o(t)$  reproduces completely the input light field with the same temporal shape and temporal order of the light pulses. Total photon number operator of the echo field signal irradiated at time $t > 2 T_o +\tau_M$:
$\hat {n}_{e} = \int_{ - \infty }^\infty {d\nu } \hat {b}_{o}^ + (\nu)
\hat {b}_{o} (\nu)  = \sum\nolimits_{k = 1}^M {\hat
{n}_{e,k} }$,
where $\hat {n}_{e,k} = \gamma _1 e^{ - 4 T_o/T_2} \int_{  \tau }^\infty {dt'} \hat
{a}_{e,k}^ + (t')\hat {a}_{e,k} (t')$ relates to the $k-th$ field mode with average photon number
$\left\langle {\hat {n}_{e,k} } \right\rangle =
e^{ - 4 T_o/T_2} Q_{e,k}\bar {n}_{1,k}$ and

\begin{equation}
\label{echo_photons}
Q_{e,k}=
\int\limits_{ - \infty }^{ + \infty } {d\nu}
[{\rm Z}_r(\nu ,\delta _{in}, T_{2}, \gamma_1, \Gamma _{r} )]^2
\frac{\left\langle {\hat {n}_{1,k} (\nu)}\right\rangle}{\bar {n}_{1,k}}.
\end{equation}

\noindent
Finally by using  Eqs. (\ref{echo-sol}), (\ref{echo_photons}) and
taking into account quantum efficiency $Q_{e,k}$ we can normalize output state $|<\psi_{out,k}(t-2T_o)|\psi_{out,k}(t-2T_o)>|\equiv 1$
for the input single photon wave packets $f_k(\nu)$ and then we obtain a fidelity $F_k=|<\psi_{in,k} (t)|\psi_{out,k}(t-2T_o)>|^2$  for the retrieved $k$-th single photon input field. The fidelity is expressed via the spectral properties of S-function and input light field as follows

\begin{equation}
\label{fidelity}
F_k=\frac{|\int_{ - \infty }^{ + \infty } {d\nu}
[{\rm S}_r(\nu ,\delta _{in}, T_{2}, \gamma_1, \Gamma _{r} )]^2|f_k(\nu)|^2|^2}
{Q_{e,k}}.
\end{equation}

\noindent
By comparing the properties of echo field $\hat b_o(t)$  with the properties of echo field emitted in the time reversal CRIB scheme in the optimal QED cavity \cite{Moiseev2010}, we see the square of S-function
$[{S}_r (\nu ,...)]^2={Z}_r (\nu ,...) \exp\{2 i \varphi (\nu) \}$
demonstrates an additional amplitude and phase modulation of the echo field (\ref{echo-sol}) while the CRIB protocol determines only  the amplitude modulation of echo field spectral components proportionally to  SS-function
${Z}_r (\nu ,\delta _{in}, T_{2}, \gamma_1, \Gamma _{r} )$.
Appearance of the additional phase modulation
(the phase $ |\varphi (\nu)|$ and imaginary part of S-function, respectively, increase with the spectral detuning $|\nu|$ )
is a result of loss of the complete temporal reversibility in the light-atom dynamics related to the echo emission in comparison with the storage stage as demonstrated recently for standard broadband AFC protocol \cite{Moiseev2012}.

\begin{figure}
\includegraphics[width=0.65\textwidth,height=0.4\textwidth]{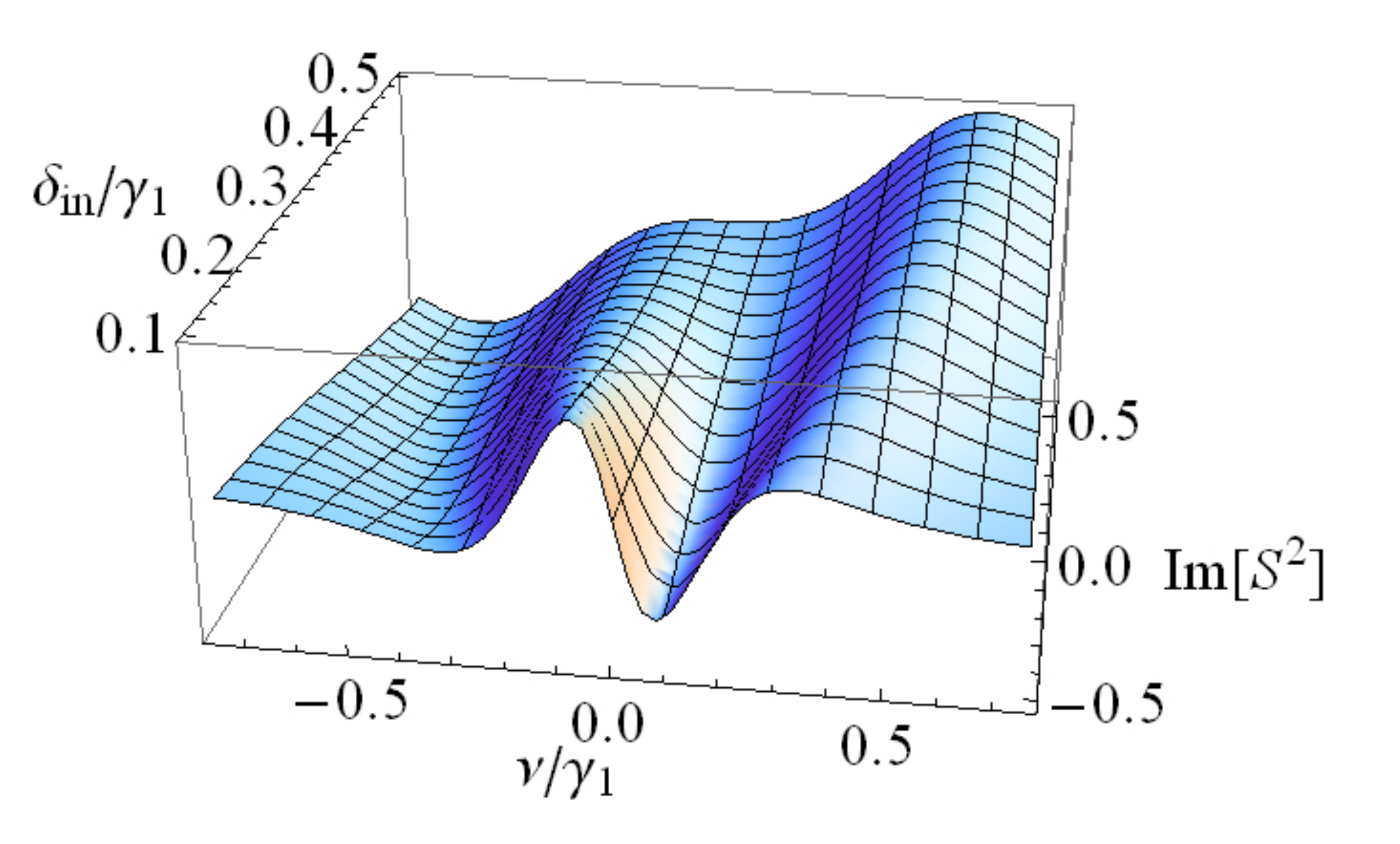}
\caption{Imaginary part of the square of S-function ($Im[S^2]$) for small inhomogeneous broadening $0.1<\delta_{in}/\gamma_1<0.5$ where almost zero phase modulation occurs within spectral range $-0.25<\nu/\gamma_1 <0.25$ if inhomogeneous broadening width $\delta_{in}\approx 0.5 \gamma_1$.}
\label{RammFig6}
\end{figure}

\begin{figure}
\includegraphics[width=0.65\textwidth,height=0.4\textwidth]{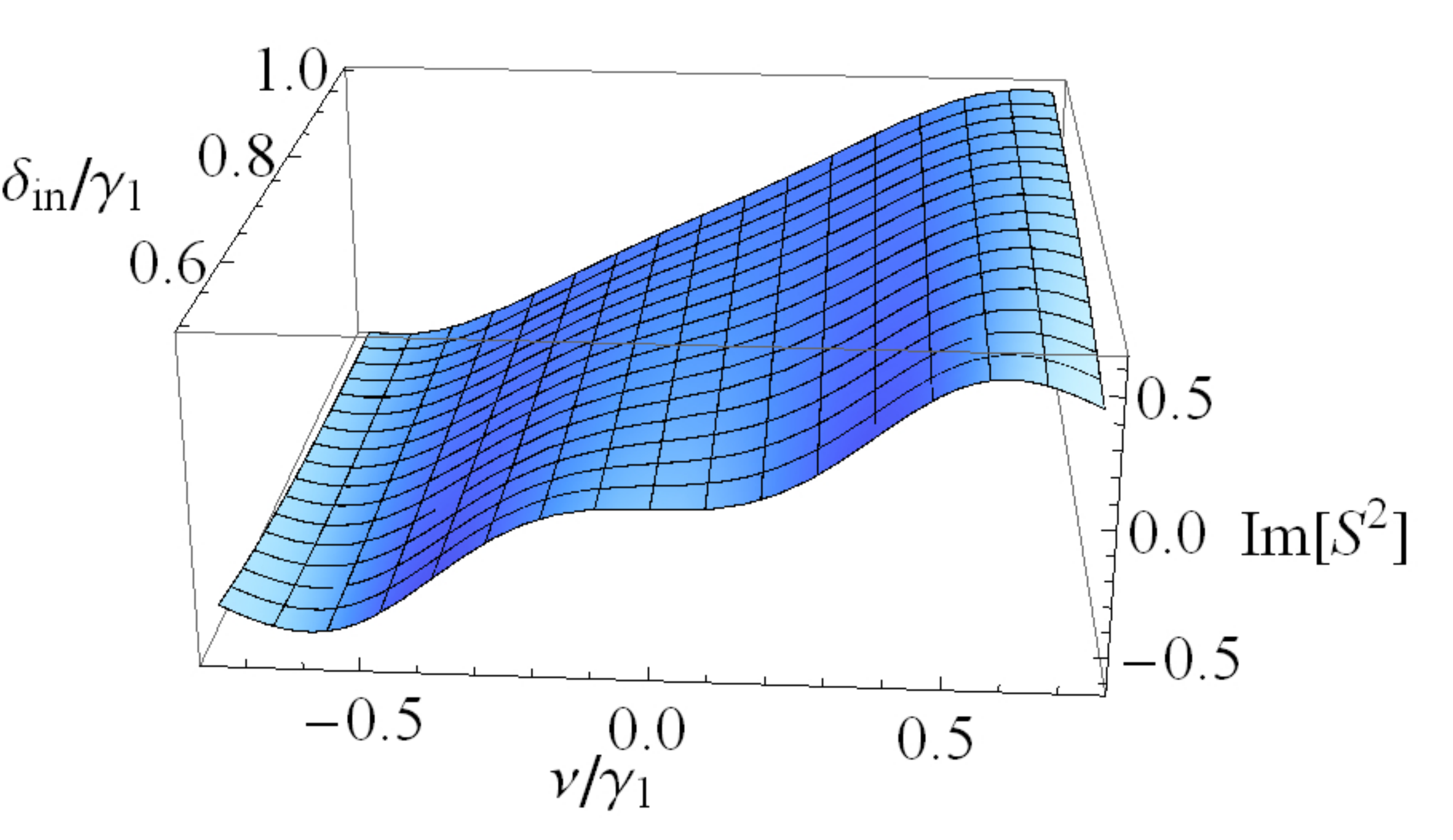}
\caption{Imaginary part of the square of S-function for larger inhomogeneous broadening $\delta_{in}$; we see almost linear phase modulation within spectral range $-0.25<\nu/\gamma_1 <0.25$  for large $\delta_{in}:$ $\delta_{in}\gg 0.5 \gamma_1$. The linear phase modulation leads only to additional time shift of each echo pulse irradiation without its temporal rephasing.}
\label{RammFig7}
\end{figure}

\begin{figure}
\includegraphics[width=0.65\textwidth,height=0.4\textwidth]{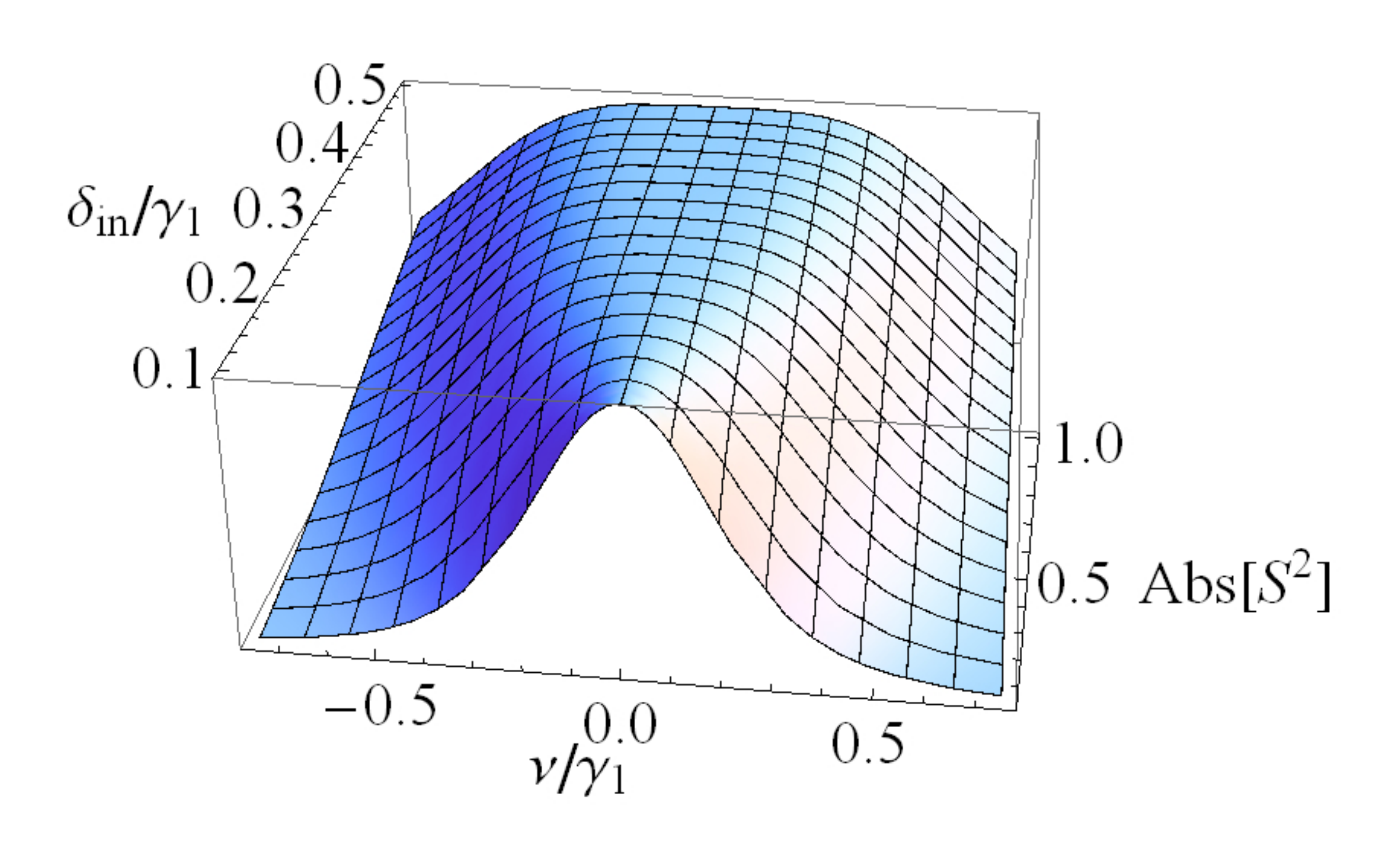}
\caption{SS-function for small inhomogeneous broadening $0.1<\delta_{in}/\gamma_1<0.5$ demonstrates perfect efficiency within spectral range $-0.25<\nu/\gamma_1 <0.25$ for inhomogeneous broadening width $\delta_{in}\approx 0.5 \gamma_1$.}
\label{RammFig8}
\end{figure}

\begin{figure}
\includegraphics[width=0.65\textwidth,height=0.4\textwidth]{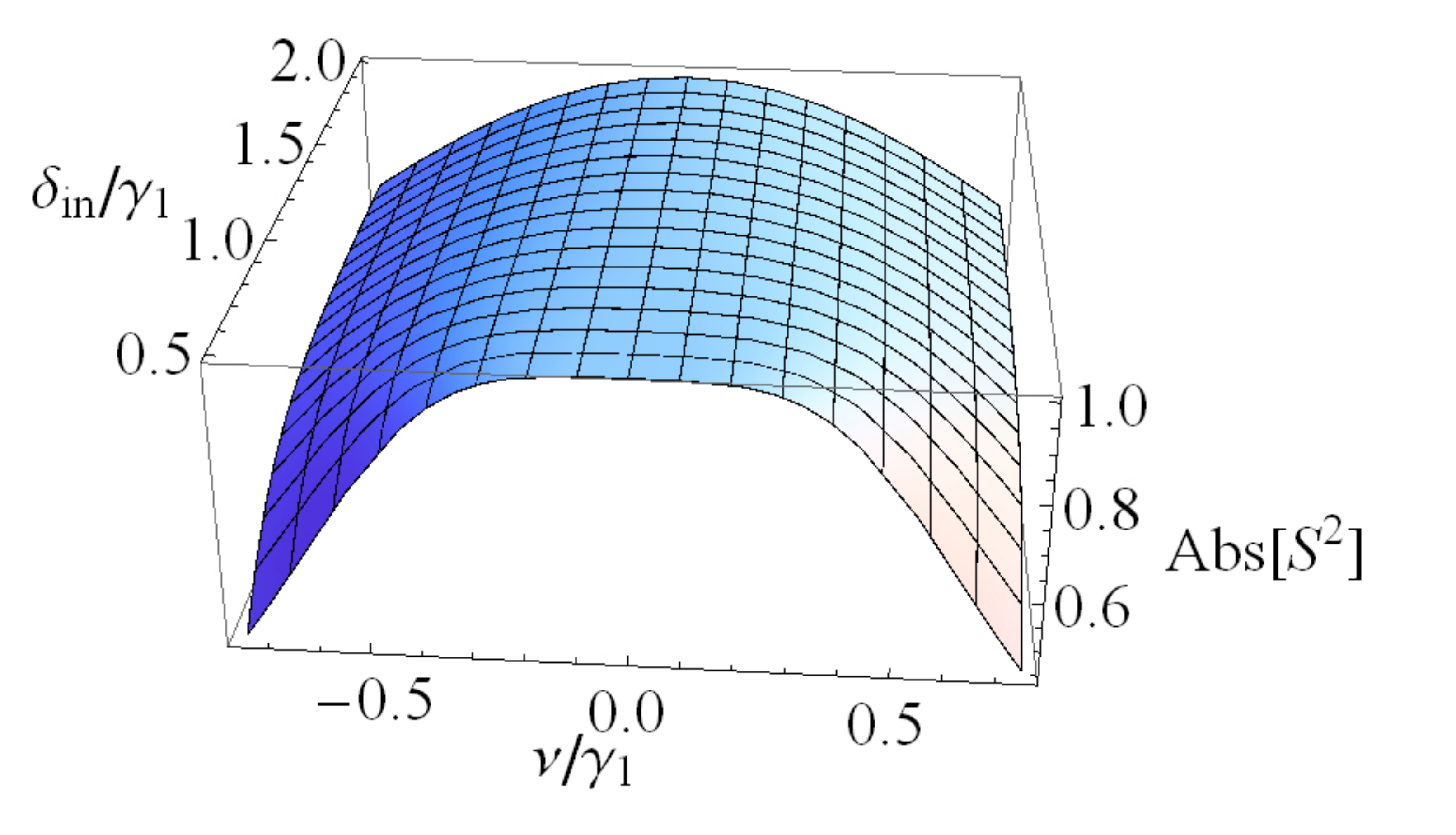}
\caption{SS-function for larger inhomogeneous broadening $\delta_{in}$; we see transfer of quantum efficiency from the flat behavior
within spectral range $-0.25<\nu/\gamma_1 <0.25$
to the quadratic spectral behavior for large inhomogeneous broadening  $\delta_{in}:$ $\delta_{in}\gg 0.5 \gamma_1$.}
\label{RammFig9}
\end{figure}

Figs. \ref{RammFig6}, \ref{RammFig7}.
show spectral properties of the imaginary part for square of S-function as function of IB width  $\delta_{in}$. It is seen that satisfying  the both matching conditions we get a
significant phase modulation only for the echo fields excited by the input signal field with larger spectral width
$\delta\omega_f>0.25\delta_{in}$.
It is important the phase modulation does not effect the quantum efficiency of echo emission in opposite to the case of AFC protocol, since the quantum efficiency of echo irradiation (\ref{echo_photons}) is determined by SS-function similar to the completely time-reversal CRIB scheme and it is determined by the small optical depth of the atomic system.
Moreover the additional phase modulations can be highly suppressed in the case of two matching conditions and narrow spectral width of the input signal fields where SS-function ${Z}_r (\nu ,...)$ and S-function ${S}_r (\nu ,...)$ will be both close to unity while the phase shift $ \varphi (\nu)$ almost vanishes in this spectral range. Spectral properties of SS-function depicted in
Figs. \ref{RammFig8}, \ref{RammFig9}. show almost ideal quantum efficiency for spectral components of input light signal in the spectral range
$-0.25\gamma_1<\nu <0.25\gamma_1$ if both matching conditions are satisfied.
Also we note that the additional phase modulation can be even completely ignored for a storage of the polarization quantum states of light (polarization photonic qubits), since both polarization components will get the same phase modulation without any influence to the self-interference and the used encoding of polarization light states.

\begin{figure}
\includegraphics[width=0.45\textwidth,height=0.3\textwidth]{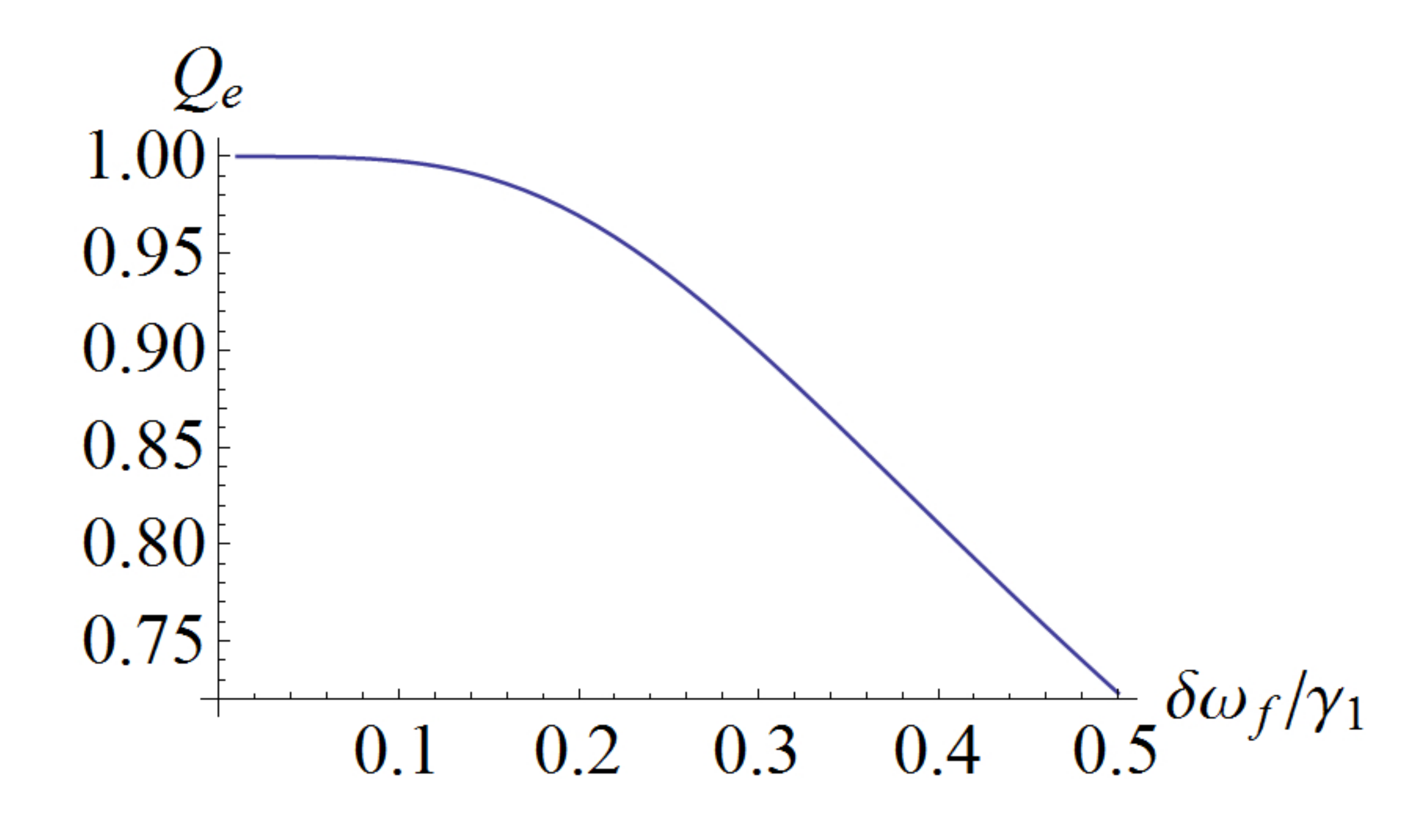}
\caption{Quantum efficiency of the echo field retrieval (\ref{echo_photons})  for input light pulse with gaussian spectral shape
$\left\langle {\hat {n}_{1,k} (\nu)}\right\rangle \sim\frac{1}{\sqrt{2\pi}\delta\omega_f} \exp \{\frac{\nu^2}{2\delta\omega_f^2}\}$ as a function of spectral width $\delta\omega_f$.}
\label{RammFig10}
\end{figure}

\begin{figure}
\includegraphics[width=0.45\textwidth,height=0.3\textwidth]{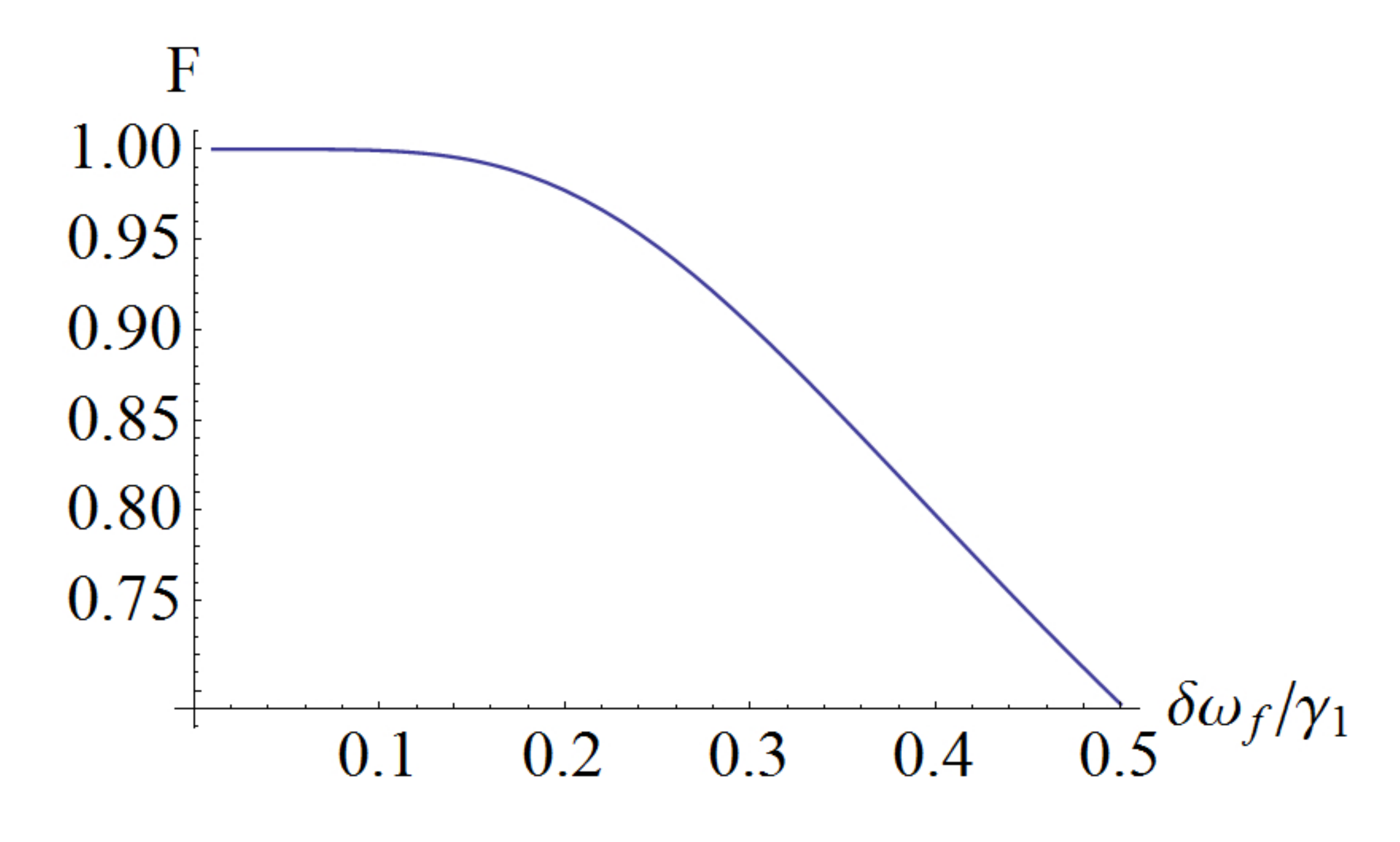}
\caption{Fidelity of the echo field retrieval (\ref{fidelity})  for input light pulse with gaussian spectral shape
$\left\langle {\hat {n}_{1,k} (\nu)}\right\rangle \sim\frac{1}{\sqrt{2\pi}\delta\omega_f} \exp \{\frac{\nu^2}{2\delta\omega_f^2}\}$ as a function of spectral width $\delta\omega_f$; fidelity decay for large spectral width of the input pulse $\delta\omega_f > 0.2 \gamma_1$ is a result of the amplitude and phase modulation of the emitted echo field.}
\label{RammFig11}
\end{figure}

Figs.\ref{RammFig10}, \ref{RammFig11}. demonstrate the quantum efficiency (\ref{echo_photons}) and fidelity (\ref{fidelity}) for retrieval of the input light pulses with gaussian spectral shape
$\left\langle {\hat {n}_{1,k} (\nu)}\right\rangle \sim\frac{1}{\sqrt{2\pi}\delta\omega_f} \exp \{\frac{\nu^2}{2\delta\omega_f^2}\}$
characterized by different spectral widths $\delta\omega_f$. As seen in the figures and in the solutions (\ref{b-eq}),(\ref{echo-sol}) we can implement a highly perfect retrieval of the input signal field with spectral width $\delta\omega_f < 0.2 \gamma_1$ when the irradiated field will be only damped by the atomic decoherence for relatively large storage time $\hat {b} (t) = \exp\{-2 T_o /T_2\} \sum\nolimits_{k = 1}^M {\hat {b}_{k} (t - 2 T_o-\tau_k)}$. Increasing of the input signal light spectral width will lead to some spectral filtration and phase modulation in the echo field spectral components  $|\nu|>0.2 \gamma_1$ resulting in appropriate decrease of the quantum efficiency and fidelity as depicted in Figs.\ref{RammFig10}, \ref{RammFig11} that can determine the optimal spectral parameters of the input signal light fields.

\section{Conclusion}

Thus a new scheme of photon echo quantum memory based on the Raman atomic transition in QED cavity has been proposed.
The scheme involves a number of unique advantages which all are critical for realization of
fine robust coherent control of the interaction between the IB atomic systems and weak single photon fields.
This scheme makes it possible to use the optically thin atomic systems with natural IB on the Raman transition which can be characterized by arbitrary narrow homogeneous isochromatic groups of atoms as it occurs for the forbidden transition
for rare-earth ions in inorganic crystals \cite{Tittel2010}, NV-centers in diamond \cite{Balasubramanian2009,Maurer2012} and in other solid state systems \cite{Tyryshkin2011,Steger2012}.
It has been shown that the used optical QED cavity provides a fine control of the atomic ensembles by external laser fields propagating without any losses along transverse path to the QED cavity mode while a strong enhancement of the atomic interaction with the cavity mode field offers an easer experimental method for highly efficient quantum storage of rather arbitrary multi-mode light fields in the optical thin medium.
Thus the scheme provides a number of experimental tools immune to many typical decoherent processes for a robust implementation of the practically realizable QM.
The proposed optical QM can be easily implemented with well-known optical technique on various atomic ensembles in a broad spectral range of the light fields.
In particular it can be also realized on the planar superconducting resonators with transverse coherent laser control of resonant atomic (spin) ensembles. This scheme is promising for the quantum repeaters based on the light-atom interface coupling microwave and optical photons and for the quantum RAM  in superconducting quantum computer. All the mentioned properties indicate a great potential for using this technique in quantum repeaters and in quantum random access memory for quantum computers.

\begin{acknowledgments}
Author would like to thank Russian Foundation for Basic Research
through grant no. 12-02-91700 for partial financial support of this work.
\end{acknowledgments}

\end{document}